
\input harvmac
\noblackbox
%
\def\ihalf{{\textstyle{i\over2}}}
\lref\flow{N. Seiberg, private communication.}
\lref\gpp{G. Gibbons, D. Page and C. Pope, Comm. Math. Phys. {\bf 127} (1990)
529.}
\lref\dkv{L. Dixon, V. Kaplunovsky and C. Vafa, ``On Four-Dimensional Gauge
Theories from
Type II Superstrings'', Nucl. Phys. {\bf B294}, (1987), 43.}
\lref\dh{L. Dixon and J. Harvey, ``String Theories in Ten Dimensions without
Spacetime Supersymmetry'' Nucl. Phys. {\bf B274}, (1986), 93.}
\lref\nati{N. Seiberg, ``Observations on the Moduli Space of Superconformal
Field Theories, ''Nucl. Phys. {\bf B303}, (1986), 288;  P. Aspinwall and D.
Morrison,  ``String
Theory on K3 Surfaces,'' preprint DUK-TH-94-68, IASSNS-HEP-94/23,
hep-th/9404151.}
\lref\duff{M. Duff, Class. Quantum Grav. {\bf 5} (1988) 189. }
\lref\kv{S. Kachru and C. Vafa,``Exact Results for $N=2$ Compactifications
of Heterotic Strings,'' hep-th/9505105. }
\lref\fhsv{S. Ferrara, J. A. Harvey, A. Strominger and
C. Vafa,``Second-Quantized Mirror Symmetry,'' hep-th/9505162}
\lref\vw{C. Vafa and E. Witten, ``Dual String Pairs With $N=1$ and
$N=2$ Supersymmetry in Four Dimensions,'' hep-th/9507050.}
\lref\sw{N. Seiberg and E. Witten, ``Electric-Magnetic Duality, Monopole
Condensation, and Confinement in $N=2$ Supersymmetric Yang-Mills Theory,''
Nucl. Phys. {\bf B426} (1994) 19, hep-th/9407087.}
\lref\howpop{P.S. Howe, G. Papadopoulos and P. West, Phys. Lett.
{\bf 339B} (1994) 219.}
\lref\towpop{G. Papadopoulos and P. Townsend, ``Compactification of
$D=11$ Supergravity on Spaces of Exceptional Holonomy,'' hep-th/9506150.}
\lref\sv{S.L. Shatashvili and C. Vafa, ``Superstrings and Manifolds of
Exceptional Holonomy,'' hep-th/9407025.}
\lref\joyce{D. D. Joyce, unpublished.}
\lref\kfiba{A. Klemm, W. Lerche and P. Mayr,``K3-Fibrations and
Heterotic-Type II String Duality,'' hep-th/9506112.}
\lref\kfibb{V. Kaplunovsky, J. Louis and S. Theisen,``Aspects of Duality in
$N=2$ String Vacua,'' hep-th/9506110.}
\lref\hs{J. A. Harvey and A. Strominger, ``The Heterotic String is a
Soliton,'' to appear in Nucl. Phys. B, hep-th/9504047.}
\lref\hitchin{N. Hitchin, ``Compact Four-Dimensional Einstein
Manifolds'', J. Diff. Geometry, {\bf 9} (1974) 435.}
\lref\schsena{J. Schwarz and A. Sen, ``Duality Symmetries of 4d Heterotic
Strings,'' Phys. Lett. {\bf B312} (1993) 105, hep-th/9305185.}
\lref\schsenb{J. Schwarz and A. Sen, ``The Type IIA Dual of the
Six-Dimensional CHL Compactification, '' hep-th/9507027.}
\lref\hullt{C. Hull and P. Townsend, ``Unity of Superstring Dualities,''
hep-th/9410167, Nucl. Phys. {\bf B438} (1995) 109.}
\lref\witt{E. Witten,``String Theory Dynamics in Various Dimensions,''
hep-th/9503124.}
\lref\aspin{P. Aspinwall, talk given at the Trieste Conference on S-Duality,
June, 1995.}
\lref\aspinpt{P. Aspinwall,  ``Enhanced Gauge Symmetries and K3 Surfaces'',
hep-th/9507012.}
\lref\sens{A. Sen, ``String-String Duality Conjecture in Six Dimensions and
Charged Solitonic Strings'', hep-th/9504027.}
\lref\lerche{W. Lerche, A.N. Schellekens and N.P. Warner,
Phys. Rep. {\bf 177} (1989) 1.}

%
%
\Title{\vbox{\baselineskip12pt
\hbox{EFI-95-46} \hbox{UCSBTH-95-22}
\hbox{hep-th/9507168}}}
{\vbox{\centerline{\bf{N=1 STRING DUALITY }}}}
{\baselineskip=12pt
\bigskip
\centerline{Jeffrey A. Harvey}
\bigskip
\centerline{\sl Enrico Fermi Institute, University of Chicago}
\centerline{\sl 5640 Ellis Avenue, Chicago, IL 60637 }
\bigskip
\centerline{David A. Lowe and Andrew Strominger}
\bigskip
\centerline{\sl Physics Department}
\centerline{\sl University of California}
\centerline{\sl Santa Barbara, CA 93206-9530}
\medskip
\bigskip
\centerline{\bf Abstract}
We discuss duality between Type IIA string theory,
eleven-dimensional supergravity,
and heterotic string
theory in four spacetime dimensions with $N=1$ supersymmetry.
We find theories whose infrared limit is trivial at enhanced
symmetry points as well as theories with $N=1$ supersymmetry but
the field content of $N=4$ theories which flow to the $N=4$ fixed
line in the infrared.
}
\Date{7/95}

%
%
\newsec{Introduction}
There has recently been dramatic progress in understanding
non-perturbative aspects of string theory. Much of this progress
centers around the idea of duality which allows one to study
phenomena in a strongly coupled theory by relating these phenomena
to weak coupling properties of a dual theory. At present the
best understood dual pairs have $N=4$ supersymmetry when reduced
to $D=4$ spacetime dimensions \refs{
\nati\hullt\schsena\witt \sens \hs \schsenb{--} \aspin}.
There is however  increasing
evidence that duality also extends to theories with $N=2$
supersymmetry \refs{\kv,\fhsv}. These theories have much richer
dynamics than $N=4$ theories as has been beautifully demonstrated
in the global case \sw. Of course realistic
chiral theories can have at most $N=1$ supersymmetry
and an understanding of the dynamics of such string compactifications
is one of the most important unsolved problems in string theory.
It is thus natural to ask whether string duality might
be extended to theories with $N=1$ supersymmetry and whether
this duality can be used to study the dynamics of such theories.
In this paper we will provide a partial answer to this question.
We will construct two different types of dual pairs with $N=1$
supersymmetry but we will see that both have rather simple
dynamics at low-energies. The construction of more realistic
$N=1$ dual pairs remains an important open problem.

In this paper we will utilize the dictionary provided by
the soliton string construction of \sens\ and \hs\ and
the orbifold techniques
developed in \fhsv\ to construct dual $N=1$ pairs. We first
review some of the essential elements of \fhsv\ and discuss
the specific symmetries we will utilize in our orbifold
constructions. We then
construct two such dual pairs, one involving an asymmetric
orbifold compactification of the IIA string with $(2,1)$ world
sheet supersymmetry which is dual to a
heterotic compactification on a Calabi-Yau
manifold and the other a duality between
a compactification of
eleven-dimensional supergravity (or strongly coupled
IIA string theory) on a seven-manifold of $G_2$
holonomy and a Calabi-Yau compactification of the heterotic string.
We will argue that both these pairs of theories preserve supersymmetry
non-perturbatively and do not generate a superpotential. In the first
example this is because there are no low-energy gauge groups except
at enhanced symmetry points and the gauge groups at enhanced symmetry points
are not asymptotically free. In the second example this is due to the fact that
the low-energy theory has $N=4$ field content at enhanced symmetry points
much as in the construction of \fhsv. We end with some brief conclusions.

\newsec{Orbifolds and Duality}
Hull and Townsend \hullt\ conjectured the existence of a duality relating
the IIA theory compactified
on $K3$ at weak coupling to the heterotic string on a four-torus
at strong coupling. After compactification of both theories on
a two-torus down to four dimensions we obtain a dual pair with
$N=4$ spacetime supersymmetry. We can then try to obtain further
dual pairs by twisting. To do so we must understand how symmetries
of one theory map onto symmetries of the other.

Let us start with
the IIA theory on $K3 \times T^2$. We can twist this theory by
a geometrical symmetry of $K3 \times T^2$. Such a symmetry acts
on the cohomology of $K3$. In particular it acts on the $22$ elements
of $H^2(K3)$. The three self-dual elements of $H^2(K3)$ gives rise
to three right-moving heterotic string coordinates while the nineteen
anti-self-dual elements of $H^2(K3)$ yield left-moving heterotic string
coordinates \hs. Combining these with the coordinates on $T^2$ which are
common to both sides a geometrical action on $K3 \times T^2$ gives
rise to an action on $(21,5)$ (left,right) moving coordinates of
the heterotic string.

The extra $(1,1)$ (left,right) coordinate of the heterotic string, $X_0$,
arises as a zero mode of the $U(1)$ Ramond-Ramond (RR)
field in ten dimensions
and is independent of $K3$. If we are to obtain $N=1$ supersymmetry
on the heterotic side however we must twist all six internal
right-moving coordinates with the twists lying in a $SU(3)$ subgroup
of $SO(6)$ (but not in an $SU(2)$ or smaller subgroup if we are
to obtain only $N=1$ supersymmetry). We are thus faced with an immediate
problem since geometrical symmetries do not seem to act on $X_0$.
We will consider two solutions to this problem. First, we can consider
non-geometrical symmetries of the IIA theory. In particular, the IIA
theory has a symmetry called $(-1)^{F_L}$ which acts as $-1$ on all
states which have fermions arising from the left-moving sector of the
theory \refs{\dkv, \dh}. In particular, $(-1)^{F_L}$ is $-1$
acting on all states in
the RR sector. Since the $(20,4)$ coordinates of the heterotic
string on $T^4$ all arise from the RR sector of the theory $(-1)^{F_L}$
maps to a symmetry which is $-1$ acting on all these coordinates, including
$X_0$. Our second solution involves recalling that the $U(1)$ RR gauge
field of the IIA theory arises via the Kaluza-Klein mechanism from
compactification of eleven-dimensional supergravity down to ten
dimensions on an $S^1$. If we denote the $S^1$ coordinate by $X_{11}$
then the geometrical action $X_{11} \to - X_{11}$ changes the sign
of the $U(1)$ gauge field and thus induces the transformation
$X_0 \to - X_0$ on the heterotic side. In the context of
compactifications of d=11 supergravity one could also consider more general
twists which mix $X_{11}$ with the other compactified coordinates
but we will not do so in this paper.

Since $K3$ plays a central role in duality
it is reasonable to expect that symmetries of $K3$ surfaces will
also play a central role in orbifold extensions of duality.
At the moment the precise rules for constructing dual pairs via
orbifolds are not understood. In particular, it is clear that
there are subtleties associated with orbifolds constructed from
symmetries which do not act freely (although these have been
understood in special cases \schsenb ). As a result we will restrict
our attention in this paper to freely acting symmetries of $K3$
surfaces. It is known that $K3$ has at most a $Z_2 \times Z_2$
group of freely acting symmetries \hitchin. The first of these which we will
call $E$ can
be taken to be the Enriques involution discussed in \fhsv. The
second which we call $A$ is an anti-holomorphic involution.
A construction of $E$ and $A$ for a class of $K3$ surfaces was
found using algebraic geometry in \hitchin.
In a particular $T^4/Z_2$ orbifold limit of $K3$ we can construct
$E$ and $A$ as follows. Let $(z_1,z_2)$ be complex coordinates
on a $T^4$ defined by the periodic identifications $z_i \sim z_1 +1$,
$z_i\sim z_i + i$ and define the $Z_2$ transformations
\eqn\one{\eqalign{\Theta: & \quad (z_1,z_2) \to (-z_1,-z_2) \cr
                   E: & \quad (z_1,z_2) \to (-z_1+ \half,z_2+ \half ) \cr
                   A: & \quad (z_1,z_2) \to (\bar z_1+ \half, \bar z_2
                                             + \ihalf ) ~.\cr }}
Then dividing by the action of $\Theta$ gives an orbifold limit of
$K3$ on which $E$ and $A$ act freely. Note that in real coordinates
$E$ and $A$ have identical actions up to relabeling of coordinates
but there is no basis in which both $E$ and $A$ act holomorphically.
In the following sections we will use $\Theta$, $E$, and $A$ with some
modifications to break the spacetime supersymmetry from $N=8$ to
$N=4$, $N=2$ and $N=1$ respectively.

\newsec{Type IIA - Heterotic duality}
We can obtain a $N=1$ compactification of the type IIA theory by
proceeding in stages, first breaking $N=4$ to $N=2$ and then $N=2$
to $N=1$. To break to $N=2$ we start with the construction in \fhsv\
of a Calabi-Yau space
\eqn\two{X = { K3 \times T^2 \over Z_2^E }~,}
where $Z_2^E$ acts as the Enriques involution $E$ on $K3$ and as
an involution on $T^2$. In the orbifold limit of $K3$ described
earlier and with $z_3$ a complex coordinate on $T^2$ we can write
this action as
\eqn\three{Z_2^E: \quad (z_1,z_2,z_3) \to (-z_1+ \half,z_2+ \half, -z_3 )~, }
plus a RR gauge transformation which has no effect on perturbative
IIA states but which is required for modular invariance on the heterotic
side.
Note that $Z_2^E$ preserves the holomorphic three form $dz_1 \wedge
dz_2 \wedge dz_3$ so $X$
is Calabi-Yau.

Let us recall the mapping of this action to
the heterotic side discussed in \fhsv. We decompose the Narain lattice
for a six-dimensional toroidal compactification as
$\Gamma^{22,6} = \Gamma^{(20,4)} \oplus \Gamma^{(2,2)} $ with
the first factor associated with the original four-dimensional toroidal
compactification and the second factor to the additional $T^2$. We denote
an element of this lattice by $|p,q \rangle$ with $p \in \Gamma^{(20,4)}$
and $q \in \Gamma^{(2,2)}$. In order to obtain a lattice
compatible with a $K3$ surface having an action of $E$ we further decompose
the first factor as
\eqn\threea{ \Gamma^{(20,4)} = \Gamma^{(9,1)} \oplus \Gamma^{(9,1)}
                       \oplus \Gamma^{(1,1)} \oplus \Gamma^{(1,1)} ~.}
%
The action of $Z_2^E$ on the heterotic side is given
by interchange of the first two factors in \threea, $-1$ on the third
factor, and a $Z_2$ shift in the fourth factor. As discussed in \fhsv,
in the twisted sector the left and right-moving vacuum energies differ
by $1/4$. Level matching thus requires a shift as described with the
shift vector $\delta$ having length squared $\delta^2 = 1/2$. This
shift corresponds to a RR gauge transformation in the dual IIA theory.

\subsec{Twisting by $(-1)^{F_L}$}

Now let us discuss the action of
$(-1)^{F_L}$.\footnote{$^*$}{The models of
this and the following subsection were
independently found and developed in somewhat more detail by
Vafa and Witten \vw .} In the IIA theory a twist
just by $(-1)^{F_L}$ simply takes the IIA theory to the IIB theory. This
is because the twist kills the $(R,NS)$ and $(R,R)$ sectors in the IIA
theory but then adds them back in with the opposite chirality for
spinors arising from the left in the twisted sector. We
can however obtain a non-trivial twist by combining the action of $(-1)^{F_L}$
with an order two shift on $T^2$. At a generic radius there are no massless
states in the twisted sector and the supersymmetry is thus reduced by half.

On the heterotic side $(-1)^{F_L}$ maps to a twist which
acts as $|p,q \rangle \to | -p,q \rangle $. This twist has
$(20,4)$ eigenvalues $-1$ on the (left,right). In the twisted sector
the vacuum energies are $E_L=1/4$ and $E_R=0$. Thus, as in the previous
example,
the twist must be accompanied by a shift with $\delta^2=1/2$ in order to
maintain
modular invariance. Since only the lattice
$\Gamma^{(2,2)}$ is left invariant we must put the shift in this factor.

Before combining these two actions let us first consider the dual pair of
theories we obtain by modding out by the generalized action of
$(-1)^{F_L}$ on both the heterotic and Type IIA sides.
On the heterotic side we have seen the
twist has $20$ eigenvalues of $-1$  acting on the left and four
eigenvalues $-1$ on the right.
As a result the low-energy theory has $20$ massless hypermultiplets
and $4$ massless vector multiplets arising from the $4$ $U(1)$ gauge
fields from $T^2$. At generic points in the moduli space this is
the full low-energy gauge theory. This spectrum agrees
with a similar analysis on the IIA side.

We can get larger gauge symmetry by going to enhanced
symmetry points in the $\Gamma^{(2,2)}$
lattice. For example, we can go to a point with $SU(2) \times SU(2)$
symmetry by going to the self-dual radius in each $S^1$ of
$\Gamma^{(2,2)}$. The projection by the shift vector leaves the
adjoint representation invariant so in the low-energy theory
we find a $N=2$ theory with $SU(2) \times SU(2)$ gauge group
(in perturbation theory)
and no matter fields. Unlike the example in \fhsv, this theory
will have non-trivial quantum corrections to the vector multiplet
moduli space \sw. This is certainly allowed in the heterotic theory
since the dilaton is in a vector multiplet.

What is the interpretation
of these corrections on the IIA side? In the $N=2$ theories constructed
in \fhsv\ the dilaton was in a hypermultiplet on the IIA side and
such spacetime quantum corrections to the vector multiplet moduli
space were forbidden on the IIA side, thus allowing a purely classical
computation of the vector multiplet moduli space. In the example
at hand this is no longer the case. The twist by $(-1)^{F_L}$ on the
IIA side kills all the spacetime supersymmetries coming from the left
and leaves invariant all those from the right. In worldsheet language
this yields a $(4,1)$ worldsheet theory. As in the heterotic string where
all the supersymmetry comes from right-movers, this puts the dilaton
in a vector multiplet on the IIA side as well. As a result the quantum
corrections to the vector multiplet moduli space cannot be exactly determined
by a classical computation in either theory. However the weakly coupled type
IIA theory corresponds to the strongly coupled heterotic theory. Thus we should
not
expect to see the enhanced $SU(2)$ symmetry in the IIA perturbation theory
because, once quantum effects are included, $SU(2)$ is not restored anywhere in
the
the moduli space of the pure gauge $N=2$ theory \sw. This is indeed
consistent with our
construction: there are no enhanced symmetries in the type II theory
at the self-dual radius of either $S^1$ of
$\Gamma^{(2,2)}$.

\subsec{An N=1 Example}
We can now go on to further break the $N=2$ supersymmetry down to $N=1$
by combining the action of $Z_2^E$ and $(-1)^{F_L}$. We have already
constructed these two symmetries in both theories ensuring modular
invariance on the heterotic side. This is not quite sufficient however
to ensure a consistent action on the heterotic side. To see this
consider the sector twisted by the product of the two symmetries.
The rotation part of the product acts on lattice vectors as
\eqn\threeb{ |p_1,p_2,p_3,p_4,q \rangle \to
             | -p_2,-p_1,p_3,-p_4,-q \rangle~, }
and is accompanied by shifts in the last two factors.
The $p_i$ refer to momenta in the four lattices given in
the decomposition \threea. However a shift
accompanied by a $-1$ rotation is equivalent to no shift at all as
can be seen by redefining the coordinate in question. Thus
the product acts without shifts. But since the eigenvalues of the
product are the same as those of $Z_2^E$ this is not consistent with
modular invariance. However we can easily modify this by redefining
the action of $(-1)^{F_L}$ on the heterotic side to include an additional
shift by $\delta$ in the first $\Gamma^{(1,1)}$ factor. This has no
effect in the sector twisted by $(-1)^{F_L}$ since the coordinate is
inverted but cures the problem with modular invariance in the sector
twisted by the product.
It is not hard to see that the massless spectrum of these two $N=1$
theories agrees at generic points in the moduli space.
Unfortunately
at generic points the low-energy spectrum does not include any gauge
fields. However we can find low-energy gauge theories by going to
the enhanced symmetry points discussed in \fhsv\ and then further projecting
by the action of $(-1)^{F_L}$. One finds enhanced symmetry groups
but with non-asymptotically free dynamics \vw.

We now turn to a $N=1$ example which has a different low-energy
structure including a gauge group at generic points.

\newsec{Eleven-dimensional Supergravity-Heterotic Duality}

As discussed in section 2, we can obtain the action $X_0 \to - X_0$
on the heterotic side by viewing this on the IIA side as resulting from
an inversion of the coordinate $X_{11}$ in the Kaluza-Klein reduction of
eleven-dimensional supergravity  down to the IIA theory. In this section
we will work directly with compactification of $d=11$ supergravity on
a seven-manifold to obtain $N=1$ supersymmetry in $d=4$. In order to
obtain a dual pair we can utilize the conjectured duality
\refs{\hullt, \witt} between
$d=11$ supergravity on $K3$ and a $T^3$ compactification of the heterotic
string. We can compactify both sides down to four dimensions on a further
$T^3$ and then twist in order to obtain a dual pair. In doing this
there are two important points we must keep in mind. First, this duality
is at least at the moment under less control than string-string duality.
In particular, it is not clear how one would deal with orbifolds having
fixed points in $d=11$ supergravity. Thus we will require that any
symmetries we mod out by act freely on $K3 \times T^3$. Second,
if we are to obtain only $N=1$ supersymmetry in $d=4$ then the
seven-manifold we compactify on must have $G_2$ holonomy. In general
a seven-manifold will have $SO(7)$ holonomy. The action of this on
the $8_s$ spinor representation of $SO(7)$ will break all the
supersymmetry. If we are to obtain precisely one supersymmetry then
we must choose the holonomy in a subgroup of $SO(7)$ for which the
$8_s$ has a single invariant component. This defines the embedding
of $G_2$ in $SO(7)$. Such compactifications have been discussed
previously in \refs{\gpp \howpop\towpop{--}\sv} and the construction of
manifolds of $G_2$ holonomy has been described in \joyce.

We can construct a seven-manifold satisfying the above constraints
by a generalization of the construction of the Calabi-Yau manifold
$X$. Namely, we consider the quotient
\eqn\four{Y = {X \times S^1 \over Z_2^A }~,}
where $Z_2^A$ acts as the freely acting anti-holomorphic involution
$A$ on $K3$, as the anti-holomorphic involution $z_3 \to \bar z_3$ on
the complex coordinate on the $T^2$ in the double cover of $X$, and
as $X_{11} \to - X_{11}$ on the $S^1$ coordinate. Seven-dimensional
spinors can be constructed as direct sums of positive and negative
chirality six-dimensional spinors. The Calabi-Yau space $X$ contains
two opposite chirality covariantly constant six-dimensional spinors,
$\eta_+$ and $\eta_-$, which are exchanged under complex conjugation.
The sum of these, $\eta_+ + \eta_-$, is the unique covariantly constant
seven-dimensional spinor on the quotient space $Y$. As described above,
the existence of a
single covariantly constant spinor implies that
$Y$ has $G_2$ holonomy.

In an orbifold
limit of $K3$ we can construct $Y$ as the quotient
\eqn\five{T^7 \over Z_2^\Theta \times Z_2^E \times Z_2^A ~,}
where
\eqn\six{\eqalign{Z_2^\Theta: & \quad (z_1,z_2,z_3,X_{11}) \to
                                 (-z_1,-z_2,z_3,X_{11}) \cr
            Z_2^E: & \quad (z_1,z_2,z_3,X_{11}) \to
                                (-z_1+ \half,z_2+ \half,-z_3,
                                    X_{11}+\half ) \cr
            Z_2^A: & \quad (z_1,z_2,z_3,X_{11}) \to
            (\bar z_1+ \half, \bar z_2 + \ihalf,
            \bar z_3 +\half +\ihalf, -X_{11} ) ~.\cr }}
The shift of $X_{11}$ by one half in $Z_2^E$ corresponds to the RR
gauge transformation discussed below \three. With the $z_3$ shift
included in $Z_2^A$, it is easily seen that
$Z_2^E$, $Z_2^A$ and $Z_2^E Z_2^A$ are all related
by a change of basis and all act freely on both $K3$ and $T^3$.

$Z_2^E $ and $Z_2^A$ can actually be defined away from the
orbifold limit as long as the four quadruplets of blown-up fixed points
which are interchanged by the symmetries have been blown up in an identical
manner.
In particular the quotient $Y$ can be constructed at the Aspinwall
points \aspinpt\ of $K3$ with an $SU(2)^{16}$ enhanced gauge symmetry.
The quotient will
then have $SU(2)^{4}$. In subsection 4.2 the dual heterotic theory at this
point in the moduli space will be explicitly constructed.

The Betti numbers of $Y$ are $b_1(Y)=b_6(Y)=0$, $b_2(Y)= b_5(Y)=4$
and $b_3(Y)=b_4(Y)=19$. One way to check this is as follows. Of the $22$
elements of $H^2(K3)$, $4$ have eigenvalues $(1,1)$ under $(Z_2^E,Z_2^A)$,
and $6$ each have eigenvalues $(1,-1)$, $(-1,1)$ and $(-1,-1)$. The
$3$ elements of $H^1(T^3)$ have eigenvalues $(1,-1)$, $(-1,1)$ and
$(-1,-1)$. Elements of $H^2(Y)$ arise from elements of $H^2(K3)$ or
elements of $H^2(T^3)$ which are invariant under both $Z^2_E$ and
$Z^2_A$. This gives $b_2(Y)=4$. Elements of $H^3(Y)$ arise from the
wedge product of two-forms on $K3$ and one forms on $T^3$ which are
invariant and from the volume form on $T^3$. This gives $b_3(Y)=18+1=19$.
The remaining Betti numbers follow from (Hodge) duality.

Reduction of the three-form potential in
$d=11$ supergravity on $Y$ gives $b_2$ $U(1)$ gauge fields in
$d=4$ and $b_3(Y)$ massless scalars. There are $b_3(Y)$ more
massless scalars associated with deformations of the
metric.  The low-energy theory thus consists of $N=1$ supergravity coupled to
four vector supermultiplets and nineteen massless chiral multiplets,
at least at generic points in the $Y$ moduli space.

\subsec{The Heterotic Dual}

We can now map this compactification of $d=11$ supergravity to a dual
$N=1$ compactification of the heterotic string.
The starting point for the construction of the
orbifold on the heterotic side is an even self-dual
Lorentzian lattice $\Gamma^{(22,6)}$ which admits
an appropriate $Z_2^E \times Z_2^A$ action.
Examples of such lattices are constructed in the next section
at enhanced symmetry points. A general lattice vector of such a lattice
may be written in the following form
\eqn\latvec{
|p_1, p_2, p_3, p_4, q_1, q_2, q_3, r_1,r_2,r_3 \rangle \in
\Gamma^{(19,3)} \oplus \Gamma^{(1,1)}
\oplus \Gamma^{(1,1)}\oplus \Gamma^{(1,1)}~,
}
where $p_i$ are four component left-moving vectors and $q_i$ are vectors
with one left-moving and one right-moving component, such that
$|p,q\rangle \in \Gamma^{(19,3)}$. The $r_i$ label points on the
three $\Gamma^{(1,1)}$ factors
corresponding to the torus $T^3$.
The action of $Z_2^E$ is
\eqn\eact{
\eqalign{
| p_1, p_2, p_3, p_4,& q_1,q_2,q_3, r_1,r_2,r_3\rangle \to
\cr & e^{2\pi i \delta_E \cdot  r_1 }
| p_3, p_4, p_1, p_2, -q_1, -q_2, q_3, r_1, -r_2, -r_3
\rangle ~. \cr}
}
The shift satisfies $\delta_E^2 = 1/2$. This acts in the first $\Gamma^{(1,1)}$
corresponding to $X_{11}$.
Likewise, the action of $Z_2^A$ is
\eqn\aact{
\eqalign{
| p_1, p_2, p_3, p_4, &q_1,q_2,q_3,r_1,r_2,r_3\rangle \to
\cr & e^{2\pi i (\delta^{(1)}_A \cdot r_2 +
\delta^{(2)}_A \cdot r_3 )}
| p_2, p_1, p_4, p_3, q_1, -q_2, -q_3, -r_1, r_2, -r_3
\rangle ~. \cr}
}
The shift $\delta_A^{(1)}$ satisfies $\delta_A^{(1) 2} = 1/2$
and acts on the second $\Gamma^{(1,1)}$. Likewise the shift $\delta_A^{(2)}$
satisfies $\delta_A^{(2) 2} = 1/2$ and acts on the third $\Gamma^{(1,1)}$.
$Z_2^E$, $Z_2^A$ and $Z_2^E \times Z_2^A$ then act freely. All these shifts are
fixed by demanding level matching in the twisted sectors.

At a generic point on the lattice, the massless
spectrum consists of four $U(1)$ vector supermultiplets
together with nineteen complex chiral supermultiplets.
Working in the RNS formalism, the four vector multiplets
arise from the four left-moving bosonic states invariant
under the $Z_2^E \times Z_2^A$  twists, combined with
the two invariant bosonic right-moving vacuum states.
Massless scalars arise from
$\alpha_{-1}^I |0\rangle_L \otimes | i \rangle_R$
($I=1,\cdots, 22,~i=1,\cdots, 6$)
projected onto invariant states.
Six of these chiral multiplets correspond to deformations of the
Calabi-Yau geometry. Twelve of the chiral multiplets
arise from the chiral multiplet components of the four
$N=4$, $d=4$ vector multiplets which survive the
projection onto invariant states. The remaining states
give rise to the gravitational multiplet and a single
chiral multiplet containing the dilaton. There
are generically no massless states in the twisted sectors,
for the choices of shifts described above.

\subsec{Enhanced Symmetry Points}

We now construct heterotic string theories compactified on
$T^3$ at a point of enhanced gauge symmetry
which admit an action of $Z_2^E \times Z_2^A$ as described
in the previous section. These will be dual to $d=11$ supergravity
compactified on a certain degenerate $K3$ surfaces.
The starting point is the even, self-dual
Lorentzian lattice $\Gamma^{(19,3)}$ of the form
\eqn\stlat{
\Lambda=\Gamma^8 \oplus \Gamma^8 \oplus \Gamma^{(1,1)}\oplus \Gamma^{(1,1)}
\oplus \Gamma^{(1,1)}~,
}
where $\Gamma^8$ is the root lattice of $E_8$.
An even self-dual lattice with the desired properties may be obtained
by orbifolding this lattice by a series of shifts. Further
details of this general procedure may be found in \lerche.

We first shift by the vector
$\delta = (1, 0^7;1, 0^7; \half, 0^2)(\half, 0,0)$, where the
first bracket denotes a shift acting on the left, the
second the shift acting on the right. Exponents
denote repeated entries. This shift satisfies
level-matching and generates a lattice which breaks the $E_8 \times E_8$
gauge group down to $SO(16) \times SO(16)$. To construct the lattice
generated by this orbifold we proceed as follows. The lattice
\stlat\ can be decomposed as a $D_8 \times D_8 \times (D_1\times D_1)^3 $
lattice with conjugacy classes added as follows. We denote the
singlet, vector, spinor and conjugate spinor conjugacy classes of $D_n$ as
$0$,$v$,$s$ and $c$ respectively.
Each $\Gamma^{(1,1)}$ corresponds to a $D_1 \times D_1$ factor
with conjugacy classes $(0,0)+(v,v) +(s,s) + (c,c)$, while
each $\Gamma^8$ gives a $D_8$ factor with conjugacy classes $(0)$ and $(s)$.
This lattice is then projected onto points $p\in \Lambda$
invariant under the action of $P = \exp(2\pi i p\cdot \delta)$
to yield the lattice $\Lambda_0$.
The even self-dual lattice generated by the shift is then
\eqn\pplat{
\Lambda'=\Lambda_0 \cup (\delta+ \Lambda_0)~.
}

A further shift by
$\delta' = ( (\half)^4 , 0^4;(\half)^4 , 0^4; 0,\half,0) (0, \half,0 )$
breaks the gauge group to $SO(8)^4$. To see this we decompose $\Lambda_0$
into a lattice with $D_4^4 \times (D_1\times D_1)^3$ symmetry plus
appropriate conjugacy classes. This lattice is then projected onto
points invariant under $P' = \exp(2\pi i p\cdot \delta')$ to yield the
lattice $\Lambda_0'$.
The final even self-dual lattice generated by this pair of shifts
is then
\eqn\ppplat{
\Lambda'' = \Lambda_0' \cup (\delta+ \Lambda_0') \cup
(\delta'+ \Lambda_0')\cup (\delta+\delta'+ \Lambda_0') ~.
}
It may be checked that this lattice admits the $Z_2^E \times Z_2^A$ symmetry.
Compactifying on a further $T^3$ and orbifolding with respect to
this symmetry, as described in the
preceding section, will yield a four-dimensional $N=1$ heterotic
theory with gauge group $SO(8)$. The associated
worldsheet currents arise as invariant linear combinations of
four level 1 currents on the original Narain lattice and so are at level 4.
As in \fhsv\ each $Z_2$  leaves invariant an $N=2$ hypermultiplet in the
adjoint of $SO(8)\times SO(8) $ arising from $Z_2$ odd combinations of
gauge currents combined with a $Z_2$ odd right-moving current.
The combination of the two $Z_2$s leaves invariant three $N=1$ chiral
multiplets  (two from an $N=2$ hypermultiplet and one from
the lower spin components of the $N=2$ vector multiplet) in the
adjoint of $SO(8)$.  Altogether this comprises the field content
of an $N=4$ vector multiplet. (Of course the moduli and
gravitational fields are in $N=1$ representations.) The three chiral multiplets
are permuted under the change of bases which permute the three non-trivial
elements of  $Z_2^A \times Z_2^E$, and there is a corresponding
global $SU(3)$ flavor symmetry.

A lattice with $SU(2)^{16}$ symmetry may be obtained by
shifting $\Lambda''$ with
$\delta'' = ( (\half)^2 , 0^2, (\half)^2 , 0^2;
(\half)^2 , 0^2, (\half)^2 , 0^2; 0^2, \half) (0^2,\half)$. Now
we decompose the lattice as $(D_2)^8 \times (D_1\times D_1)^3$,
plus conjugacy classes. As before we project onto points invariant
under the shift, and add in twisted sectors. The resulting lattice
admits the $Z_2^E \times Z_2^A$ symmetry. Upon further
compactification and orbifolding as described above, we obtain
a $N=1$ heterotic theory in four dimensions with gauge group $SU(2)^4$
at level 4 and with $N=4$ field content.

The two constructions described above give rise to rank 4 level 4
Kac-Moody algebras with $G=SO(8)$ and $G=SU(2)^4$. We presume that
other rank 4 groups such as $SU(5)$ or $SU(3)\times SU(2) \times U(1)$
can also be obtained. Consistency of our picture requires that
supersymmetry remains unbroken non-perturbatively. This is consistent
with the fact that the field content of the low-energy gauge theory
forms a finite $N=4$
representation with a global
$SU(3)$ symmetry. In the infrared this flows to a scale invariant
$N=4$ theory \flow\ which of course does not break supersymmetry on its own.

\newsec{Conclusions}

We have found examples of dual pairs of theories
with $N=1$ supersymmetry in four dimensions. We feel that
this work along with earlier constructions \refs{\kv, \fhsv}
provides convincing evidence
that string duality can be extended in a non-trivial way to
backgrounds with $N=2$ and $N=1$ supersymmetry. However the
$N=1$ dual pairs constructed so far are rather simple and
have non-generic low-energy behavior. In more realistic $N=1$
theories we would expect to find spontaneous supersymmetry
breaking, perhaps through gluino condensation. In analogy with
the work of \kv\ and \refs{\kfibb,\kfiba} it may be that
more interesting pairs can be
found by compactification of $d=11$
supergravity on seven-manifolds of $G_2$ holonomy constructed as
$K3$ fibrations of three-manifolds.

\bigskip
{\bf Acknowledgements}

While this work was in progress we received eprints \vw\ and
\towpop\ which
have substantial overlap with our work.
We acknowledge conversations with D. Kutasov. We would
like to thank
the International Center for Theoretical Physics at Trieste and the Aspen
Center for Physics for providing a stimulating atmosphere for the beginning
and completion of this work respectively. This work was supported in
part by NSF Grants PHY 91-23780, PHY 91-16964 and DOE Grant No. DOE-91ER40618.

\listrefs
\bye